# Edge states in polariton honeycomb lattices

M. Milićević[1], T. Ozawa[2], P. Andreakou[1], I. Carusotto[2], T. Jacqmin[1,*], E. Galopin[1], A. Lemaître[1], L. Le Gratiet[1], I. Sagnes[1], J. Bloch[1], A. Amo[1]

[1]*Laboratoire de Photonique et Nanostructures, LPN/CNRS, Route de Nozay, 91460 Marcoussis, France*

[2]*INO-CNR BEC Center and Dipartimento di Fisica, Università di Trento, I-38123, Povo, Italy*

**The experimental study of edge states in atomically-thin layered materials remains a challenge due to the difficult control of the geometry of the sample terminations, the stability of dangling bonds and the need to measure local properties. In the case of graphene, localised edge modes have been predicted in zig-zag and bearded edges, characterised by flat dispersions connecting the Dirac points. Polaritons in semiconductor microcavities have recently emerged as an extraordinary photonic platform to emulate 1D and 2D Hamiltonians, allowing the direct visualization of the wavefunctions in both real- and momentum-space as well as of the energy dispersion of eigenstates via photoluminescence experiments. Here we report on the observation of edge states in a honeycomb lattice of coupled micropillars. The lowest two bands of this structure arise from the coupling of the lowest energy modes of the micropillars, and emulate the π and π\* bands of graphene. We show the momentum space dispersion of the edge states associated to the zig-zag and bearded edges, holding unidimensional quasi-flat bands. Additionally, we evaluate polarisation effects characteristic of polaritons on the properties of these states.**

**Introduction**

Graphene is a two-dimensional material with extraordinary transport properties. Many of them arise from its non-trivial geometry with two identical atoms per unit cell, resulting in linear bands crossing at two non-equivalent Dirac points. The spinor character of the wavefunctions gives rise to a Berry phase of $\pm\pi$ when circumventing each of these points in momentum space. This feature is at the origin of its non-conventional transport properties like ballistic Klein propagation [1,2], antilocalisation in the presence of disorder [3], or Veselago lensing effects when traversing a potential step [4]. The non-zero Berry phase around the Dirac points has an interesting consequence: the existence of edge states in finite size samples. Indeed, it has been recently shown that the existence of such states can be related to the non-zero Berry phase along a straight trajectory in momentum space defined by the geometry of the considered edge [5–7]. Because the Berry phase depends on the trajectory, not all possible edge geometries present localised states [8].

The most commonly considered graphene terminations are the so-called armchair, zig-zag and bearded. Among them, only the last two present localised states, characterised by a flat dispersion linking the K and K' Dirac points [8–11]. Though these edge states may play an important role in the localisation and transport in small size graphene nanoribbons, experimental studies on the spatial distributions of the wavefunctions and their dispersion is not straightforward. While different kinds of terminations can be prepared in graphene and be visualised by scanning tunnelling microscopy [12,13], the existence of electronic edge states has only been evidenced via the measurement of the local density of states, which provides information on their energy and on the curvature of their dispersion, but misses any information on their microscopic spatial structure and on their momentum distribution [12].

Photonic graphene analogues are an ideal platform to experimentally address the single particle physics of two-dimensional lattices [14]. Optically-induced honeycomb lattices in photorefractive crystals have been employed to study conical diffraction effects [15,16] and the spinor character of the honeycomb lattice [17]. Arrays of photonic coupled waveguides can be engineered with single site precision, and they have been recently used to engineer artificial gauge fields in strained honeycomb lattices [18] and to fabricate a photonic analogue of a Floquet-Chern insulator [19]. Lattices of microwave resonators have also been shown to mimic several properties of electronic graphene [20,21]. The possibility to control both the local geometry and the coupling has been used in both systems to study exciting phenomena like the topological transition associated to the merging of Dirac cones [7,22], as first suggested by Montambaux and co-workers [23,24], and edge states. Moreover, photonic systems allow realizing any type of lattice termination, even those which are not stable in graphene such as the bearded edge. The spatial and momentum distributions of certain edge wavefunctions have been studied using microwave resonators [25] and coupled waveguides [7,26]. However, neither of these two systems provides the combined information on real, momentum and energy spaces needed to reconstruct the band dispersion of the eigenfunctions, and in particular of the edge states.

In this sense, arrays of coupled micropillars in semiconductor microcavities provide a versatile platform to study one- and two-dimensional photonic lattices. In a single micropillar, photons are confined in the three spatial dimensions, and they are strongly coupled to quantum well excitons placed at the maxima of the electromagnetic field. The new eigenstates of the micropillars are polaritons, with a mixed exciton-photon nature that provides them with significant interactions [27]. By partially overlapping two micropillars, we can engineer the hopping of photons, and thus polaritons, between different pillars [28,29]. By extending this coupling to two-dimensional arrays, a polariton honeycomb lattice has been recently realised [30]. Other techniques to engineer polariton lattices have been recently reported [31–36].

The coupled micropillar system is well described by a tight binding Hamiltonian giving rise to polariton dispersions analogue to the electronic $\pi$ and $\pi^*$ bands of graphene. One of its main assets is that the escape of photons out of the microcavity provides all the information regarding the amplitude, phase, momentum and energy of the polariton eigenstates: angularly resolved spectroscopy reveals the energy bands of the system, evidencing the characteristic linear dispersion around the Dirac cones, as shown in Ref. [30]. In the present work, we report on the observation of localised edge states along zig-zag and bearded edges in such a honeycomb lattice of coupled micropillars. We observe a flat-band dispersion for these edge states, connecting K and K' points at complementary regions in momentum space, as expected from tight-binding calculations [8]. Despite the non-zero next-to-nearest neighbour coupling in our lattices, the observed edge states remain flat up to the resolution given by the polariton linewidth. Our results are promising in view of observing topologically protected edge states when combining polariton polarisation effects and external magnetic fields to realise a photonic topological insulator [37,38].

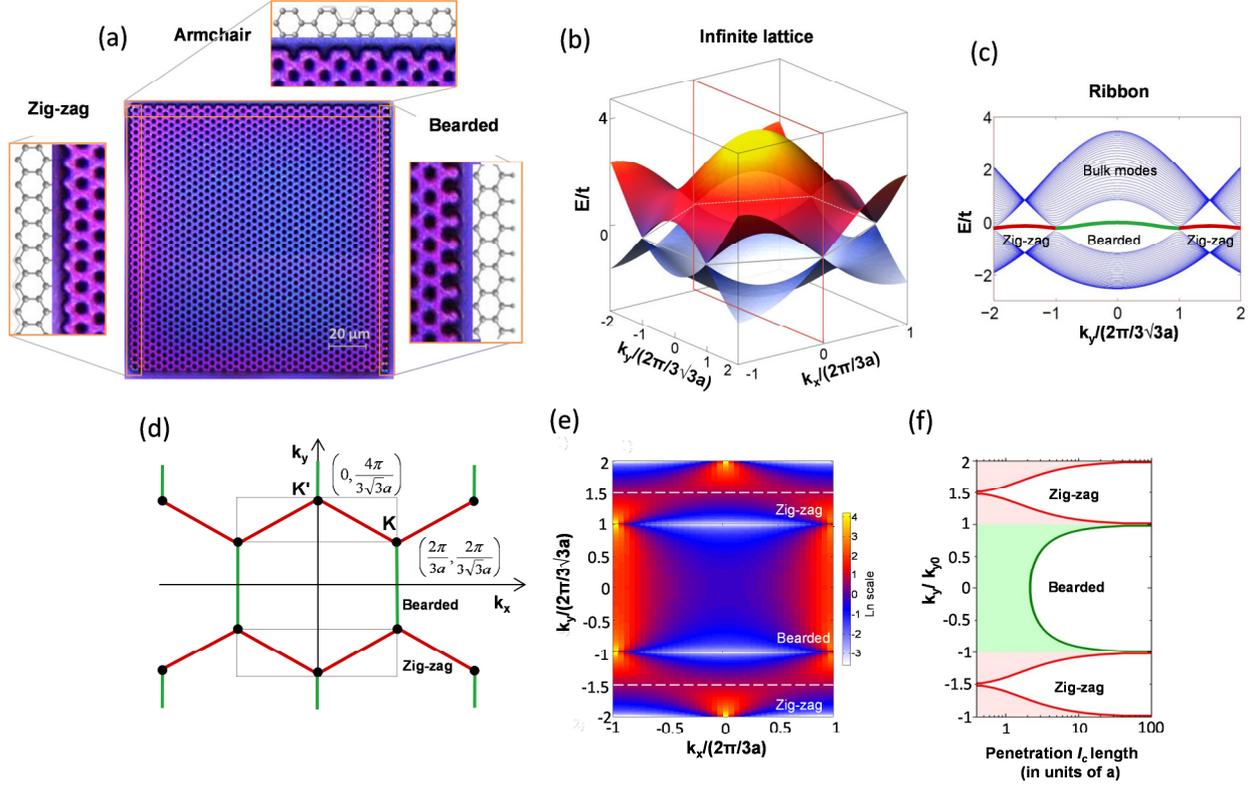

**Figure 1. Honeycomb lattice edges**. (a) Optical microscope image of the sample containing the three considered types of edges. (b) Calculated band-structure of an infinite honeycomb lattice in the tight-binding approximation with nearest- and next-nearest-neighbour coupling. (c) Calculated band-structure for graphene nanoribbons with bearded (green) and zig-zag (red) edges. The different blue lines correspond to the projection on the $k_y$-E plane of the dispersion of the different transverse modes due to the confinement in the x-direction. The red and green lines show the edge bands corresponding to zig-zag and bearded terminations, respectively. (d) First and adjacent Brillouin zones showing the regions in k space where the edge states are expected. (e) Simulation of the momentum distribution for zig-zag and bearded edge states obtained by Fourier transforming along x the calculated spatial wavefunction of the edge states corresponding to different $k_y$ values. Dashed lines show fully delocalised edge states along $k_x$. (f) Penetration length of the amplitude of the edge states wavefunction according to equation (1).

## The polariton honeycomb lattice

In our experiments we use a $Ga_{0.05}Al_{0.95}As$ $\lambda/2$ cavity embedded in two $Ga_{0.05}Al_{0.95}As/Ga_{0.8}Al_{0.2}As$ Bragg mirrors with 28 (40) top (bottom) pairs. The cavity contains three sets of four 70Å GaAs quantum wells located at the three central maxima of the confined electromagnetic field, resulting in a Rabi splitting of 15meV. The planar microcavity, grown by molecular beam epitaxy, is etched down to the substrate in the form of a series of honeycomb lattices of coupled micropillars. The zero dimensionality of the micropillars imposes quantized energy levels for polaritons. Therefore, they behave like artificial photonic atoms. The lowest energy polariton eigenstate of an individual micropillar presents cylindrical symmetry, like the $p_z$ orbitals of graphene. To introduce the coupling between the micropillars, we etch them such that they partially overlap (the interpillar distance is set to be smaller than their diameter). The narrow region between the pillars represents a potential barrier for photons and thus, for polaritons, through which they can evanescently tunnel. The coupling strength can be tuned by choosing the size of the pillars and the distance between them [28]. To enhance the tunnelling we consider lattices with predominantly photonic polaritons, at -17meV exciton-photon detuning.

By properly designing the lithographic mask used to etch the planar cavity into a honeycomb lattice, we engineer different types of edges in our samples. Figure 1(a) shows a lattice containing the most commonly considered edge types: zig-zag, armchair and bearded. The lattice consists of nearly 30 unit cells along the crystallographic axes. This size is large enough for the properties of the bulk to be dominant when probing lattice sites located near the centre, while simultaneously showing edge physics when probing the properties in the edges.

Before reporting on the experimental results, we first consider the graphene dispersion relation for the bulk and edge bands using a tight binding model including next-nearest-neighbour hopping $t'$=0.08 $t$, where $t$ is the nearest neighbour coupling. This is the value used in Ref. [30] to describe our lattices. Figure 1(b) shows the calculated momentum-energy relation for an infinite honeycomb lattice without edges. It features positive and negative energy bands intersecting at six Dirac points in the first Brillouin zone. To calculate the dispersion of the edge states we consider a nanoribbon geometry: an infinite lattice in the $y$ direction and of finite width in the $x$ direction, ending with the same type of boundary on both sides. Therefore, the calculated dispersions are continuous along $k_y$, with several transverse modes corresponding to the confinement in the $x$ direction. The result is shown in figure 1(c) for ribbons with either zig-zag or bearded edges. Each of the different transverse modes corresponds to each individual line in the figures. Edge bands appear for the zig-zag and bearded edges in complementary regions of $k_y$, connecting the Dirac cones [8,9], as indicated in red and green, respectively in figure 1(d): the zig-zag edge band appears for $k_{y(zig\text{-}zag)}$ ∈ [-2,-1]$k_{y0}$ ∪ [1, 2]$k_{y0}$, and the bearded edge band for $k_{y(bearded)}$ ∈ [-1, 1]$k_{y0}$, with $k_{y0}$ = 2π/(3√3$a$) and $a$ being the interpillar distance. The dispersion of the edge states deviates from a perfect flatband as a consequence of the next-nearest-neighbour hopping parameter being included in the calculation. However this deviation is rather small: 50 μeV in total for a value of $t$=250 μeV.

Spatially, the edge states are localised on the outermost sites, with an exponentially decaying amplitude into the bulk ($\psi_{edge}(x) \sim e^{-x/l_e}$). In the absence of next-nearest-neighbour coupling the penetration length follows [8]:

$$l_e = \frac{3a}{2\left|ln\left(2\,cos(k_y\sqrt{3}a/2)\right)\right|} \qquad (1)$$

The finite penetration results in a finite width in momentum space for the edge states. Figure 1(e) shows the $k_x$-$k_y$ momentum distribution of the zig-zag and bearded edge states calculated by Fourier transforming with respect to $x$ the spatial distributions of the edge state for each $k_y$ as obtained from the solution of the tight-binding Hamiltonian. The edge modes are spread around straight lines connecting the Dirac points at the border of the Brillouin zone, as schematically represented in figure 1(d). The edge states with $k_y$ corresponding to the centre of the zig-zag band ($k_y$=±1.5 $k_{y0}$) are fully delocalised in the $k_y$ direction (see dashed line in figure 1(e)). Correspondingly, these states are spatially fully localised, down to a single site (see figure 1(f)). In the case of the bearded edge state, maximum spatial localisation is attained at $k_y$ =0, with a penetration length of 2.2$a$, larger than the maximally localised zig-zag edge state. At the Dirac points the penetration length becomes infinite, and the edge states merge into bulk modes. Note that no edge state is formed for armchair edges.

To experimentally access the polariton wavefunctions and dispersions we perform low temperature (10K) photoluminescence experiments. We excite the sample non-resonantly using a Ti:Sapph monomode laser at 740 nm, about 100 meV above the lowest band of the honeycomb lattice. The excitation creates electron–hole pairs in the quantum wells, which relax incoherently and, under low power excitation, populate all polaritonic energy bands. We analyse the far field emission arising from photons escaping out of the cavity. Owing to momentum-conservation laws, each photon is emitted with in-plane momentum equal to the in-plane momentum of the polariton in which it originated. Thus, there is direct correspondence between the angle of emission and the in-plane momentum of polaritons up to a reciprocal lattice vector. Each angle of emission corresponds to a point in the Fourier plane of the collecting lens, a high numerical aperture microscope objective (NA = 0.65), which is also used for the excitation. By imaging the Fourier plane on the entrance slit of a spectrometer, we resolve in energy and in-plane momentum the far-field emission along the line given by the slit (parallel to $k_y$), for a given value of $k_x$, which we record on a CCD camera. By varying the position of the image of the Fourier plane on the slit we collect the dispersion for different values of $k_x$. We are thus able to reconstruct a 3D matrix whose axis are $k_x$, $k_y$ and the emission energy [39]. The described tomography process is also carried out for the real-space emission to reconstruct the spatial distribution of the emitted light at a given energy, and study the localisation of the edge state. We select the linear polarization of the emission using a set of half-waveplates and linear polarisers.

We study a graphene simulator similar to the one shown on figure 1(a), containing zig-zag edges. The diameter of the pillars ($d$ = 3 μm) and the interpillar distance ($a$ = 2.4μm) result in a significant tunnelling strength, $t$=250 μeV, in combination with a relatively narrow linewidth ~150 μeV. For the excitation we focus the laser in a Gaussian spot with a diameter of 3 μm, covering around one pillar. We select the emission linearly polarised along the $y$ axis, parallel to the edge. Since the emission arises mainly from the excited area we are able to selectively image the dispersion from the bulk or the edge. Figure 2(a) shows the momentum space emission at the energy of the Dirac point (zero energy) when exciting the lattice in the bulk. We observe six isolated bright spots at the Dirac points, which identify the first Brillouin zone hexagon. These are the points in which upper and lower bands meet (figure 1(b)). The triangular shape of the points is due to the trigonal warping known to be present when next-nearest-neighbour tunnelling is present. Figure 2(b) shows the energy resolved far field emission along line 1, parallel to $k_y$ at $k_x$ = 1.7·(2π/3$a$). We select a line passing through the second Brillouin zone in order to evidence the upper band, whose emission is strongly reduced in the first Brillouin zone due to destructive interference effects [30]. In figure 2(b) we can identify the upper and lower energy bands separated by a gap as expected for the graphene dispersion for this value of $k_x$ (figure 1(b)). The black curve in figure 2(b) depicts the dispersion expected from the tight-binding approximation with $t$=250 μeV and $t'$=0.08t (i.e., 20 μeV). Note that the next-nearest neighbour coupling is evidenced via the asymmetry of the bands above and below $E_0$.

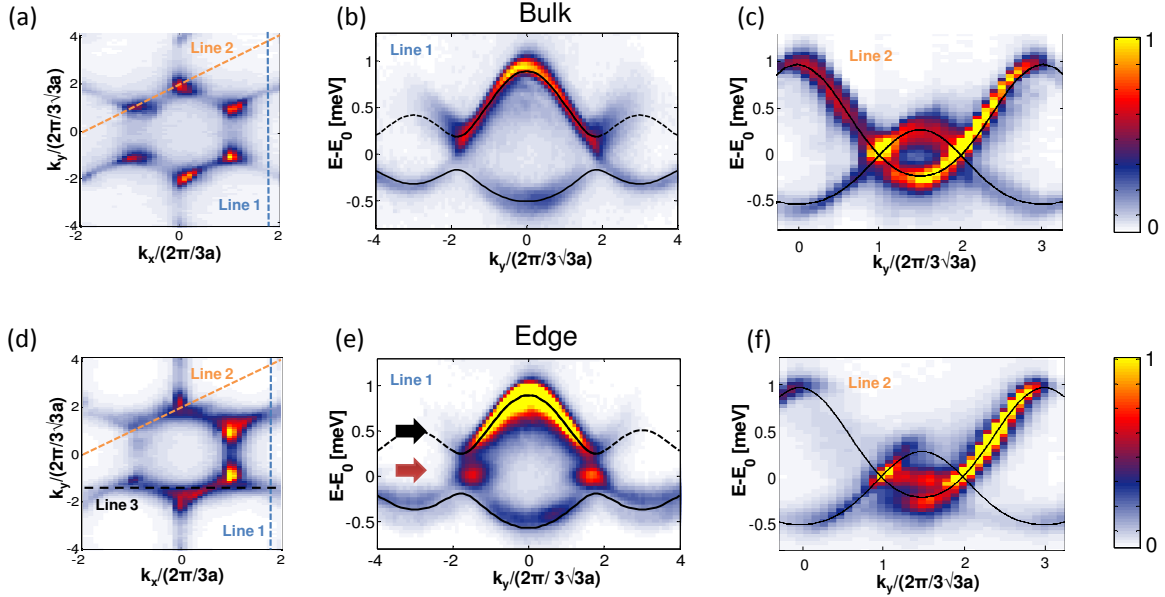

**Figure 2. Zig-zag edge, momentum space emission.** (a),(d) Measured photoluminescence intensity in momentum space at the energy of the Dirac points ($E_0$=1569.2 meV) under bulk (a) and zig-zag edge (d) excitation. (b), (e) Spectrally resolved far-field emission along line 1 in (a) and (d), passing through the second Brillouin zone for excitation in the bulk (b) and in the zig-zag edge (d). (c), (f) Measured dispersion along line 2 in (b) and (d), respectively. The black lines show fits to the tight-binding honeycomb dispersion.

**Zig-zag edge**

We now address the situation when the excitation spot is moved to one of the external pillars forming the zig-zag edge. Figure 2(d) shows the luminescence at the energy of the Dirac points for this excitation configuration. The Dirac cones are now continuously connected by a bright line in the $k_{y(zig-zag)}$ region while there is a dark region in the middle at the $k_{y(bearded)}$ region, as expected from figure 1(d,e). Additionally, along line 3 we observe a spread emission in $k_x$, indicating a fully localized edge mode. This feature matches the state marked by a dashed line in the simulation shown in figure 1(e). The overall emitted intensity in momentum space is asymmetric since light is collected at the edge and translational symmetry is broken. When analysing the energy resolved emission along line 1 (figure 2(e)), two additional lobes are clearly observed in the gap between the upper and lower bands. Their location in momentum space corresponds to that expected for the edge states shown in figure 1(d) (red lines). The measured full width at half maximum of the lobes along the $k_y$ direction in figure 2(e) is 0.75$k_{y0}$, in agreement with the theoretical prediction for the edge states along the same line in momentum space extracted from the simulation shown in figure 1(e), within a 20% error.

The quasi-dispersionless character of the band associated to the edge states can be evidenced by selecting a spectral cut along line 2 in figure 2(d), which contains the $k_{y(zig-zag)}$ region. Figure 2(f) shows a flat band linking the two Dirac cones. No such state is present in the bulk (figure 2(c)), where only the corresponding bulk dispersion is detected. Only the states with group velocities propagating towards the bulk (positive slope) emit light, explaining the asymmetry of figure 2(f). For a clearer comparison with the edge states band, a fit of the bulk bands is presented in figure 2(e),(f) by a black curve. Although our system exhibits effects of next-nearest-neighbour tunnelling for the bulk bands, the edge states band stays flat within the linewidth. Indeed the magnitude of the curvature obtained in the tight-binding calculations (50 µeV), is small compared to the emission linewidth (150 µeV). Note that emission from bulk states is also present in figure 2(e)-(f).

In addition to momentum space imaging, our system allows evidencing the localization of the edge states by looking at the real space emission. For this purpose we use large Gaussian laser spot, 20µm in diameter, covering around 30 pillars. In this way we are able to excite edge modes on several pillars, and to compare the emission of the edge and bulk states from a single set of measurements. Figure 3(a) shows the emitted intensity at the energy of the middle of the upper bulk band, 0.5 meV above the Dirac points (black arrow in figure 2(e)). The bulk modes present the expected honeycomb pattern, with an intensity distribution following the pump spot. Figure 3(c) shows the real space emission at the energy of the edge state ($E_0$, red arrow in figure 2(e)). In this case the outermost line of pillars shows a stronger emission, corresponding to the localised edge state.

This interpretation is supported by simulations of a driven-dissipative model of the honeycomb lattice. In the simulations, we added to the tight binding Hamiltonian a monochromatic resonant pump and cavity losses of $\gamma=0.1t$ for all lattice sites. We calculate the steady state with a pumping beam at $E_0$ which covers the whole sample with an incident momentum $\boldsymbol{k} = (1/(2\sqrt{3}), 3/2)k_{y0}$, corresponding to the centre of the segment connecting the Dirac points where the zig-zag edge state is expected. The result is shown in figure 3(d), revealing the edge state fully localized on the outermost pillars, as expected from equation (1). The same simulation at the energy of the bulk bands shows emission from the whole lattice, as depicted in figure 3(b).

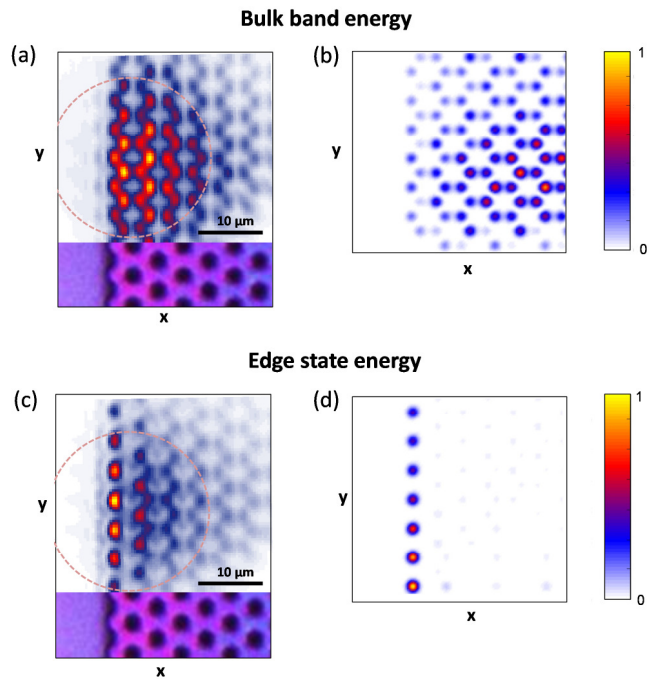

**Figure 3. Zig-Zag edge, real space emission.** (a), (c) Measured real space emission at the energy of the bulk band (a) (energy marked with a black arrow in figure 2(e)), and at the energy of the edge state (c) (red arrow in figure 2(e)). Dashed lines show the half maximum intensity of the excitation laser spot. The lower part of the panels shows an optical microscope image of the edge. (b), (d) Simulations of emission of a driven-dissipative polaritonic honeycomb lattice coherently pumped at an energy corresponding to bulk states (b), and at the energy and momentum of a zig-zag edge state (d).

One of the specific characteristics of polaritons, different from other photonic simulators like coupled waveguides or microwave resonators, is their significant polarization dependent properties. The polarization dependent penetration of the electromagnetic field in the Bragg mirrors forming the cavity result in a linearly polarized TE-TM splitting whose magnitude increases quadratically with the in-plane momentum [40], resulting in the so-called optical spin-Hall effect [41–43]. Additionally, the polarization-dependent hopping between coupled micropillars [44] has been shown to give rise to spin-orbit coupling effects in hexagonal photonic molecules in the polariton condensation regime [45]. When analysing the spontaneous emission from the bulk of the honeycomb lattice presented here in different linear polarization components, we observe negligible effects. The reason is that the period of the lattice is big enough to restrict the first Brillouin zone to small values of in-plane momenta where the TE-TM splitting is expected to be smaller than the measured linewidth. Nevertheless, we do observe significant polarization effects when analysing the emission from the edge states. Figure 4(a) reproduces figure 2(e) showing the energy resolved far field emission upon small spot excitation located at one of the outermost pillars of the zig-zag edge. Here, we select the emission linearly polarized parallel to the direction of the edge ($y$), as in all the results we have presented so far. When selecting the opposite linear polarization direction, perpendicular the edge, we observe that the edge state is located at a lower energy $\Delta E$=160 µeV. Similar polarization splittings have been reported in 1D polariton microwires [46,47]. The splitting may arise from the interplay between two effects. First, the asymmetric photonic confinement along and perpendicular to the edge could induce a linear polarization splitting on the confined photonic modes. Second, the finite size etched structure may give rise to strain crystal fields resulting in the splitting of the excitonic modes with polarization directions along and perpendicular to the strain field. In the considered structure, a strain mismatch between $x$ and $y$ directions could take place close to the edge of the honeycomb lattice. Given the significant value of $\Delta E$, the excitonic origin of the splitting seems the most likely. Indeed, photonic confinement effects are expected to result in polarization splittings of 5-10 µeV in this kind of structures [45], much smaller than the linewidth. Note that the strain field might penetrate a few sites into the lattice, thus affecting the energy of the bulk bands close to the edge. This is the origin of the observed redshift of the bulk bands in figure 4(b) with respect to figure 4(a).

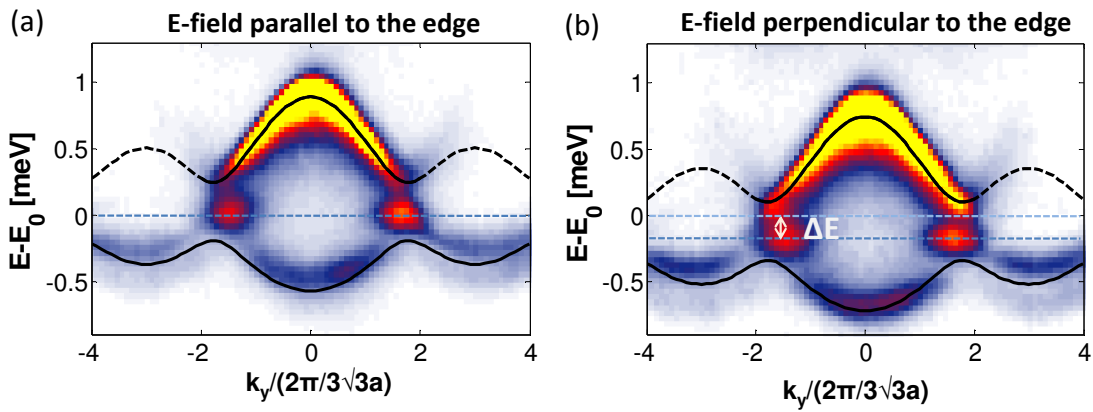

**Figure 4 Polarization effects.** Measured dispersion along line 1 in figure 2(d) when excitation is performed on the zig-zag edge. Linear polarization of detection is perpendicular (a) and parallel (b) to the edge. $\Delta E$ indicates the energy splitting between the edge modes with opposite linear polarizations.

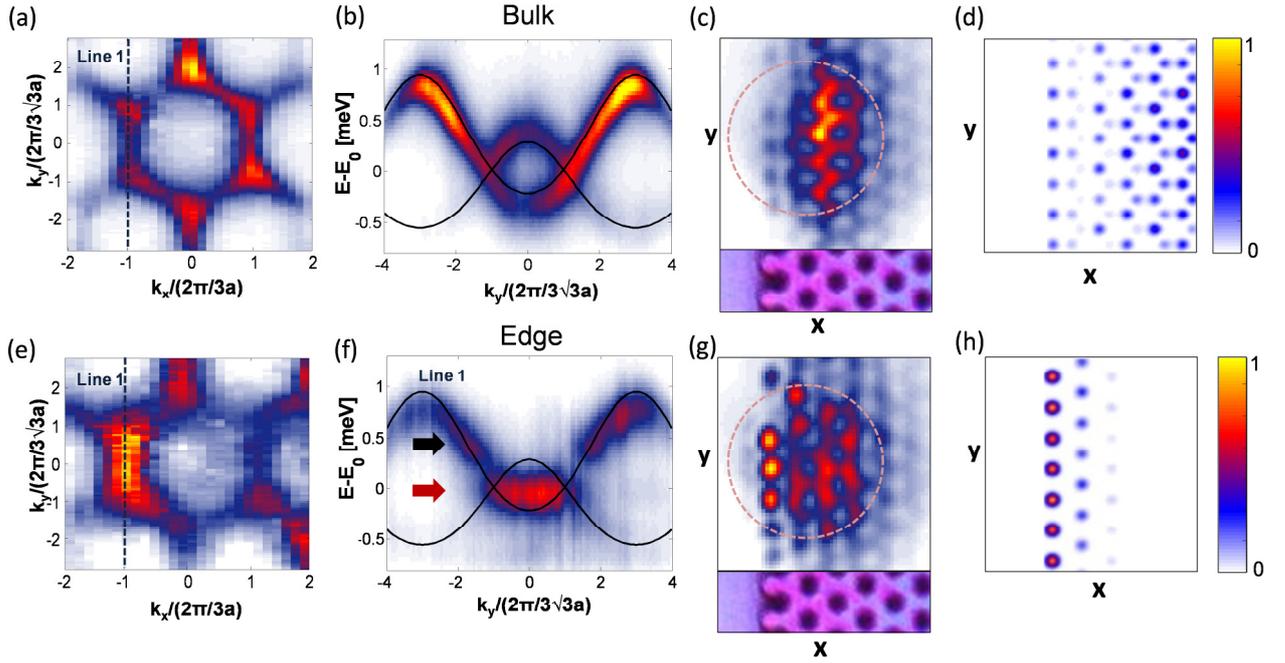

**Figure 5. Bearded edge.** (a), (e) Measured photoluminescence intensity in momentum space at the energy of the Dirac points, in the bulk (a) and on the bearded edge (e). (b), (f) Spectrally resolved far-field emission along line 1 in (a), passing through the region where edge states are expected, in the bulk (b) and on the bearded edge (f). (c), (g) Measured real space emission at the energy of the bulk band (black arrow in (f)), and at the energy of the edge state (red arrow in (f)). $E_0$=1578.1 meV. Dashed lines show the half maximum intensity of the excitation laser spot. The polarization of detection is parallel to the edge (vertical). (d), (h) Simulations corresponding to resonant excitation of bulk and edge modes, respectively.

**Bearded edge**

Bearded edges have also been predicted to exhibit edge states [11]. Experimental investigation of this type of edge band is not feasible in carbon graphene where dangling bonds specific to this kind of termination are chemically unstable. Thus, it has been studied mostly theoretically and using graphene analogues [8,25,26]. To study the energy-momentum dispersion of this kind of edge states, we have fabricated a lattice containing bearded edges, with pillar diameter of $d$ = 2.5 μm, and interpillar distance $a$ = 1.76 μm, giving the same tight binding tunnelling amplitudes as in the lattice with the zig-zag edges. However, the smaller pillar diameter results in non-radiative losses that give rise to a larger linewidth (~350μeV). Experiments are performed under the same conditions as described above, in both real and reciprocal space configurations. Figure 5(a) shows the momentum space at the energy of the Dirac points when exciting the bulk of the lattice. Again, we are able to identify the six Dirac points of the first Brillouin zone with gaps between them. They are less pronounced than in figure 2 due to the broader linewidth. The bulk dispersion along line 1 defined in figure 5(a), containing the $k_{y(bearded)}$ region, is shown in figure 5(b). The expected shape of the bands is observed, with crossings at two Dirac points. When the probe is placed on the edge of the sample, different patterns are observed. The momentum-space map at the Dirac point energy shows an enhanced emission in the $k_{y(bearded)}$ and equivalent regions (figure 5(e)), revealing the edge states band. Its full width at half maximum along the $k_x$ direction at $k_y=0$ is $0.50k_{x0}$, with $k_{x0} = 2\pi/(3a)$, in excellent agreement with the prediction in figure 1(e), where this value is $0.45k_{x0}$. The dispersion along line 1 (figure 5(f)) shows now a flatband connecting the two Dirac points in the momentum space region corresponding to $k_{y(bearded)}$. As previously described for the zig-zag edges, the linewidth detain us from observing non-flatness of the edge band.

To study the spatial location of the state we perform measurements and simulations of the real space emission under excitation with a large pump spot. Figure 5(h) shows a simulation of the emitted intensity when exciting the edge state at $k_y=0$. The observed bearded edge state resides on the sublattice corresponding to the bearded ending, and it penetrates several lattice sites into the bulk, as expected from equation (1). In the experiment (figure 5(g)) we observe bright spots on the outermost pillars of the lattice. This emission is absent at the energy of the bulk modes (figure 5(c)), and thus it corresponds to the edge state. The penetration depth is, however, difficult to estimate experimentally due to the emission from the bulk modes at the same energy.

**Conclusion**

We have used a photonic graphene simulator to directly visualise the localised states associated with the bearded and zig-zag types of graphene edges. Clear identification of the different kinds of edge states is possible thanks to real space and far field imaging. Although we mainly used the photonic nature of polaritons in the present experiments in a honeycomb lattice, their excitonic content offers the exciting possibility of exploring nonlinear effects [27]. Virtually unfeasible in natural graphene, phenomena such as soliton solutions to the nonlinear Dirac equation expected for instance in the armchair edge [48] can be experimentally addressed.

This work was supported by the ANR project "Quandyde" (Grant No. ANR-11-BS10-328 001), by the French RENATECH network, EU-FET Proactive grant AQuS, Project No. 640800, the ERC grants Honeypol and QGBE, and the Autonomous Province of Trento, partly under the call "Grandi Progetti 2012," project "On silicon chip quantum optics for quantum computing and secure communications—SiQuro."